\documentclass{emulateapj}
\usepackage{natbib}

\newcount\longrefs
\def\aap{\ifnum\longrefs=1 {Astron.\ Astrophys.}\else 
                           {A\hbox{\rm \&}A}\fi}
\def\aapr{\ifnum\longrefs=1 {Astron.\ Astrophys.\ Rev.}\else 
                            {A\hbox{\rm \&}AR}\fi}
\def\aaps{\ifnum\longrefs=1 {Astron.\ Astrophys.\ Suppl.}\else 
                            {A\hbox{\rm \&}AS}\fi}
\def\aj{\ifnum\longrefs=1 {Astron.\ J.}\else 
                          {AJ}\fi} 
\def\ao{\ifnum\longrefs=1 {Applied Optics}\else 
                           {Appl.\ Opt.}\fi} 
\def\aspcs{\ifnum\longrefs=1 {Astron.\ Soc.\ Pacific Conf. Series}\else 
                           {ASP Conf.\ Ser.}\fi} 
\def\apj{\ifnum\longrefs=1 {Astrophys.\ J.}\else 
                           {ApJ}\fi} 
\def\apjl{\ifnum\longrefs=1 {Astrophys.\ J. Lett.}\else 
                            {ApJ}\fi} 
\def\aplett{\ifnum\longrefs=1 {Astrophys.\ J. Lett.}\else 
                            {ApJ}\fi} 
\def\apjs{\ifnum\longrefs=1 {Astrophys.\ J. Suppl.}\else 
                            {ApJS}\fi}
\def\apss{\ifnum\longrefs=1 {Astrophys.\ and Space Science}\else 
                            {Ap\hbox{\rm \&}SS}\fi}
\def\araa{\ifnum\longrefs=1 {Ann.\ Rev.\ Astron.\ Astrophys.}\else 
                            {ARA\hbox{\rm \&}A}\fi}
\def\azh{\ifnum\longrefs=1 {Astronomicheskii Zhurnal}\else 
                            {Astron.\ Zhur.}\fi}
\def\baas{\ifnum\longrefs=1 {Bull.\ Am.\ Astron.\ Soc.}\else 
                            {BAAS}\fi}
\def\bain{\ifnum\longrefs=1 {Bull.\ Astronom.\ Institutes Netherlands}\else
                            {Bull.\ Astr.\ Inst.\ Neth.}\fi}
\def\gca{\ifnum\longrefs=1 {Geochim.\ Cosmochim.\ Acta}\else 
                           {Geochim.\ Cosmochim.\ Acta}\fi}
\def\grl{\ifnum\longrefs=1 {Geophys.\ Res.\ Lett.}\else 
                           {Geoph.\ Res.\ Lett.}\fi}
\def\iaucirc{\ifnum\longrefs=1 {IAU Circulars}\else 
                          {IAU Circ.}\fi}
\def\ip{\ifnum\longrefs=1 {in press}\else 
                          {in press}\fi}
\def\jchemp{\ifnum\longrefs=1 {J.\ Chem.\ Phys.}\else 
                           {J.\ Chem.\ Phys.}\fi}  
\def\jcp{\ifnum\longrefs=1 {J.\ Chem.\ Phys.}\else 
                           {J.\ Chem.\ Phys.}\fi}  
\def\jgr{\ifnum\longrefs=1 {J.\ Geophys.\ Res.}\else 
                           {J.\ Geophys.\ Res.}\fi}  
\def\jmolspec{\ifnum\longrefs=1 {J.\ Mol.\ Spectrosc.}\else 
                           {J.\ Mol.\ Spectrosc.}\fi}  
\def\jqsrt{\ifnum\longrefs=1 {J.\ Quant.\ Spectrosc.\ Radiat.\ Transfer}\else 
                           {J.\ Quant.\ Spectrosc.\ Radiat.\ Transfer}\fi}  
\def\jrasc{\ifnum\longrefs=1 {J.\ Royal Astron.\ Soc.\ Canada}\else 
                           {JRAS Can.}\fi}  
\def\mnras{\ifnum\longrefs=1 {Mon.\ Not.\ Roy.\ Astron.\ Soc.}\else 
                             {MNRAS}\fi} 
\def\nat{\ifnum\longrefs=1 {Nature}\else 
                           {Nat}\fi}
\def\pasj{\ifnum\longrefs=1 {Pub.\ Astron.\ Soc.\ Japan}\else 
                            {PASJ}\fi} 
\def\pasp{\ifnum\longrefs=1 {Pub.\ Astron.\ Soc.\ Pacific}\else 
                            {PASP}\fi} 
\def\physscr{\ifnum\longrefs=1 {Physica Scripta}\else 
                            {Phys.\ Scrip.}\fi} 
\def\planss{\ifnum\longrefs=1 {Planetary \& Space Science}\else 
                            {Plan. \& Space Sci.}\fi} 
\def\procspie{\ifnum\longrefs=1 {Proc.\ SPIE}\else 
                            {Proc.\ SPIE}\fi} 
\def\qjras{\ifnum\longrefs=1 {Quarterly J.\ Royal Astron.\ Soc.}\else 
                            {QJRAS}\fi} 
\def\sa{\ifnum\longrefs=1 {Soviet Astron..}\else 
                               {Sov.\ Astron.}\fi}
\def\skytel{\ifnum\longrefs=1 {Sky \& Telescope}\else 
                            {Sky \& Tel.}\fi} 
\def\solphys{\ifnum\longrefs=1 {Solar Phys.}\else 
                               {Solar Phys.}\fi}
\def\ssr{\ifnum\longrefs=1 {Space Science Rev.}\else 
                               {Space\ Sci.\ Rev.}\fi}

\def\dutch{\def\refname{Referenties}\def\abstractname{Samenvatting}%
  \def\bibname{Bibliografie}\def\chaptername{Hoofdstuk}%
  \def\appendixname{Bijlage}\def\contentsname{Inhoudsopgave}%
  \def\listfigurename{Lijst van figuren}\def\listtablename{Lijst van tabellen}%
  \def\indexname{Index}\def\figurename{Figuur}\def\tablename{Tabel}%
  \def\partname{Deel}\def\enclname{Bijlage(n)}\def\ccname{Ter attentie van}%
  \def\headtoname{Aan}\def\headpagename{Pagina}%
  \def\today{\number\day\space\ifcase\month\or januari\or februari\or maart\or%
     april\or mei\or juni\or juli\or augustus\or september\or oktober\or%
     november\or december\fi \space\number\year}%
  \typeout{
              >>>>> use hlatex209 for Dutch hyphenation <<<<< 
         }}
\newcounter{onefig} \newcounter{fignumber}
\newcount\nocaptions \newcount\nofigures \newcount\figwidth
\newcount\viewgraphs
  \def\paper{}  \def\figlabel{} 
\long\def\nextfig#1{\setcounter{figure}{\value{fignumber}}
  \addtocounter{fignumber}{1}
  \ifnum \viewgraphs=1 \newpage \pagestyle{empty} \fi 
  \ifnum\value{onefig}=0 #1 \fi                 
  \ifnum\value{onefig}=\value{fignumber} #1 \fi}
\def\figwidths#1#2{\ifnum \nocaptions=1 #2mm \else #1mm \fi}  
\def\paper#1{}  
\long\def\plotfig#1#2{\ifnum \nofigures=1 \else #2 \fi}
\long\def\captiontext#1{\ifnum \nofigures=1 \raggedright \fi 
   \ifnum \nocaptions=1 \paper
     \ifnum \viewgraphs=0 
       \newline  \mbox{}\hrulefill\mbox{} \newline 
       \newline label:~\{\figlabel\} 
     \fi 
     \else \ifnum \nofigures=0 \fi 
   #1 \fi}

\newcount\panelwidth \newcount\panelheight 
\newcount\bxmin \newcount\bymin \newcount\bxmax \newcount\bymax
\newcount\tbxmin \newcount\tbymin
\newcount\tpanelwidth \newcount\tpanelheight \newcount\tpdif
\def\panelsize #1,#2;{\panelwidth=#1 \panelheight=#2}  
\def\setbb #1,#2;#3,#4;#5,#6;{
  \tbxmin=#1 \tbymin=#2    
  \bxmin=#3 \bymin=#4      
  \bxmax=#5 \bymax=#6}     
\def\barepanel #1{%
  \ifnum\panelheight=0 
    \tpdif=\bymax \advance\tpdif by -\bymin
    \multiply \tpdif by \panelwidth
    \tpanelheight=\tpdif
    \tpdif=\bxmax \advance\tpdif by -\bxmin
    \divide \tpanelheight by \tpdif
  \else \tpanelheight=\panelheight \fi
  \epsfig{file=#1,%
     bbllx=\bxmin bp,bblly=\bymin bp,bburx=\bxmax bp,bbury=\bymax bp,clip=,%
     width=\panelwidth mm,height=\tpanelheight mm}}
\def\labelypanel #1{
  \ifnum\panelheight=0 
    \tpdif=\bymax \advance\tpdif by -\bymin
    \multiply \tpdif by \panelwidth
    \tpanelheight=\tpdif
    \tpdif=\bxmax \advance\tpdif by -\bxmin
    \divide \tpanelheight by \tpdif
  \else \tpanelheight=\panelheight \fi
  \tpdif=\bxmax \advance\tpdif by -\tbxmin
  \tpanelwidth=\panelwidth \multiply \tpanelwidth by \tpdif
  \tpdif=\bxmax \advance\tpdif by -\bxmin
  \divide \tpanelwidth by \tpdif
  \epsfig{file=#1,%
    bbllx=\tbxmin bp,bblly=\bymin bp,bburx=\bxmax bp,bbury=\bymax bp,%
    clip=,width=\tpanelwidth mm,height=\tpanelheight mm}}
\def\labelxpanel #1{%
  \ifnum\panelheight=0 
    \tpdif=\bymax \advance\tpdif by -\bymin
    \multiply \tpdif by \panelwidth
    \tpanelheight=\tpdif
    \tpdif=\bxmax \advance\tpdif by -\bxmin
    \divide \tpanelheight by \tpdif
  \else \tpanelheight=\panelheight \fi
  \tpdif=\bymax \advance\tpdif by -\tbymin
  \multiply \tpanelheight by \tpdif
  \tpdif=\bymax \advance\tpdif by -\bymin
  \divide \tpanelheight by \tpdif
  \epsfig{file=#1,%
    bbllx=\bxmin bp,bblly=\tbymin bp,bburx=\bxmax bp,bbury=\bymax bp,%
    clip=,width=\panelwidth mm,height=\tpanelheight mm}}
\def\labelxypanel #1{%
  \ifnum\panelheight=0 
    \tpdif=\bymax \advance\tpdif by -\bymin
    \multiply \tpdif by \panelwidth
    \tpanelheight=\tpdif
    \tpdif=\bxmax \advance\tpdif by -\bxmin
    \divide \tpanelheight by \tpdif
  \else \tpanelheight=\panelheight \fi
  \tpdif=\bxmax \advance\tpdif by -\tbxmin
  \tpanelwidth=\panelwidth \multiply \tpanelwidth by \tpdif
  \tpdif=\bxmax \advance\tpdif by -\bxmin
  \divide \tpanelwidth by \tpdif 
  \tpdif=\bymax \advance\tpdif by -\tbymin 
  \multiply \tpanelheight by \tpdif
  \tpdif=\bymax \advance\tpdif by -\bymin
  \divide \tpanelheight by \tpdif
  \epsfig{file=#1,%
    bbllx=\tbxmin bp,bblly=\tbymin bp,bburx=\bxmax bp,bbury=\bymax bp,%
    clip=,width=\tpanelwidth mm,height=\tpanelheight mm}}

\def\CC{\par \vspace*{-2ex} \footnotesize \baselineskip=8pt \begin{verbatim}}

\long\def\startignore #1\stopignore{}   

\def\setlistparams{         
  \topsep=0.7ex                 
  \itemsep=0.7ex                
  \leftmargini=3ex}             
\setlistparams                  

\newcounter{alistindex}       

\newcounter{romenumnr}

\newlength{\minipagewidth}

\newsavebox{\boxcontent}
\newcommand{\ovalhead}[1]{
  \unitlength=1cm
  \sbox{\boxcontent}{\mbox{~~{#1}~~}}
  \begin{center}
    \ifdim\wd\boxcontent>6ex 
    \ifdim\wd\boxcontent<8cm 
    \begin{picture}(8,3) \thicklines     
      \put(4.0,0.8){\oval(8,1.6)} 
      \put(0.0,0.7){\parbox{8cm}{
         \begin{center} \usebox{\boxcontent} \end{center}}}
    \end{picture}
    \else \ifdim\wd\boxcontent<12cm 
    \begin{picture}(12,3) \thicklines     
        \put(6.0,0.8){\oval(12,1.6)} 
        \put(0.0,0.7){\parbox{12cm}{
           \begin{center} \usebox{\boxcontent} \end{center}}}
    \end{picture}
    \else
    \begin{picture}(16,3) \thicklines     
        \put(8.0,0.8){\oval(16,1.6)} 
        \put(0.0,0.7){\parbox{16cm}{
           \begin{center} \usebox{\boxcontent} \end{center}}}
    \end{picture}
    \fi \fi \fi
  \end{center}} 

\newcounter{headnr}            
\newcounter{subheadnr}[headnr]
\newcounter{subsubheadnr}[subheadnr]
\def\head #1\par{
  \stepcounter{headnr}                          
  \vspace{2ex} \noindent                        
  {\bf \theheadnr~~~~#1}\\[1ex] \noindent}      
\def\subhead #1\par{  
  \stepcounter{subheadnr}
  \vspace{1.3ex} \noindent
  {\bf \theheadnr.\arabic{subheadnr}~~~#1}\\[0.3ex] \noindent}
\def\subsubhead #1\par{
  \stepcounter{subsubheadnr}
  \vspace{1.0ex} \noindent
  {\bf \theheadnr.\arabic{subheadnr}.\arabic{subsubheadnr}~~~#1}\\ \noindent}

\font\dropfont= cmr12 scaled \magstep5
\def\dropcap#1#2{{\noindent
    \setbox0\hbox{\dropfont #1}\setbox1\hbox{#2}\setbox2\hbox{(}%
    \count0=\ht0\advance\count0 by\dp0\count1\baselineskip
    \advance\count0 by-\ht1\advance\count0by\ht2
    \dimen1=.5ex\advance\count0by\dimen1\divide\count0 by\count1
    \advance\count0 by1\dimen0\wd0
    \advance\dimen0 by.25em\dimen1=\ht0\advance\dimen1 by-\ht1
    \global\hangindent\dimen0\global\hangafter-\count0
    \hskip-\dimen0\setbox0\hbox to\dimen0{\raise-\dimen1\box0\hss}%
    \dp0=0in\ht0=0in\box0}#2}

\def\level #1 #2#3#4{$#1 \: ^{#2} \mbox{#3} ^{#4}$}   

\def\mathstacksym#1#2#3#4#5{\def#1{\mathrel{\hbox to 0pt{\lower 
    #5\hbox{#3}\hss} \raise #4\hbox{#2}}}}

\mathstacksym\lta{$<$}{$\sim$}{1.5pt}{3.5pt} 
\mathstacksym\gta{$>$}{$\sim$}{1.5pt}{3.5pt} 
\mathstacksym\lrarrow{$\leftarrow$}{$\rightarrow$}{2pt}{1pt} 
\mathstacksym\lessgreat{$>$}{$<$}{3pt}{3pt} 

\slugcomment{Accepted for publication in the {\em{Spitzer}} edition of the Astrophysical Journal Supplement Series}

\shorttitle{Excitation near LkH$\alpha$~234}
\shortauthors{P. Morris et al.}

\begin{document}


\title{Excitation of Molecular Material Near the Young Stellar Object
LkH$\alpha$~234 in NGC~7129}


\author{Patrick~W. Morris\altaffilmark{1}, Alberto Noriega-Crespo, 
Francine R. Marleau, Harry I. Teplitz}
\affil{{\em{Spitzer}} Science Center, IPAC, Caltech, MS 
220-6, Pasadena, CA 91125}

\author{Keven I. Uchida\altaffilmark}
\affil{Center for Radiophysics and Space Research, Cornell University, Ithaca, 
NY 14853-6801}
\and 
\author{Lee Armus\altaffilmark}
\affil{{\em{Spitzer}} Science Center, IPAC, Caltech, MS 
220-6, Pasadena, CA 91125}

\altaffiltext{1}{NASA {\em{Herschel}} Science Center, IPAC, Caltech, MS 
220-6, Pasadena, CA 91125}

\setcounter{footnote}{1}

\begin{abstract}

With the {\em{Spitzer}}-IRS\footnote{The {\em{Spitzer}} Space
Telescope is operated by the Jet Propulsion Laboratory, and
Caltech Institute of Technology under NASA contract 1407. Support
for this work was provided by NASA through an award issued by
JPL/Caltech. The IRS was a collaborative venture between Cornell University and 
Ball Aerospace Corporation funded by NASA through the Jet Propulsion 
Laboratory and the Ames Research Center.} we have obtained the first mid-IR spectroscopy
of NGC~7129, in the unusually strong outflow and in a ridge of H$_2$ emission 
near the Herbig Be star LkH$\alpha$~234.  The UV radiation field strength is
estimated from PAH band intensities in the H$_2$ ridge, and found
to be comparable to that of NGC~7023.  From the rotational 
H$_2$ emission lines we have deduced aperture average excitation temperatures
and column densities in the two regions, finding the H$_2$ ridge values to
be consistent with pumping by UV fluorescence, but also comparable to warm-gas 
regions of Cep~A that form H$_2$ in non-dissociative C-shocks.   
The H$_2$ emission in the outflow is consistent with formation by 
collisional excitation in J-shocks, with shock velocities of 10 -- 30 km sec$^{-1}$.
A photodissociating component may be present in the outflow, by similarity of S(0) 
line intensities in both regions.  There is no indication of warm dust in the 
outflow. We also present the first 16 $\mu$m imaging of a Galactic nebula using the unique
imaging capabilities of the IRS, 
and combine with ground-based 2.12 $\mu$m (H$_2$ 1 -- 0 S(1)) imaging.
Candidate pre-main sequence objects are clearly evident in these data.  We also 
find extended emission not previously observed around the young B star 
BD~+65$^\circ$1638, $\sim 22''$  across, showing that the region is not free 
of material as otherwise inferred by recent high angular resolution mapping
at submillimeter wavelengths.  The presence of this material complicates
interpretation of the surrounding CO cavity and origin(s) of the photodissociated
region, and further spectroscopic observations are needed to characterize its nature. 

\end{abstract}


\keywords{infrared: ISM --- ISM: jets and outflows --- ISM: individual
(NGC~7129) --- reflection nebulae --- stars: formation
  --- stars: pre-main sequence --- infrared: stars}


\begin{figure*}[th!]
\begin{center}
\includegraphics[height=500pt,angle=270]{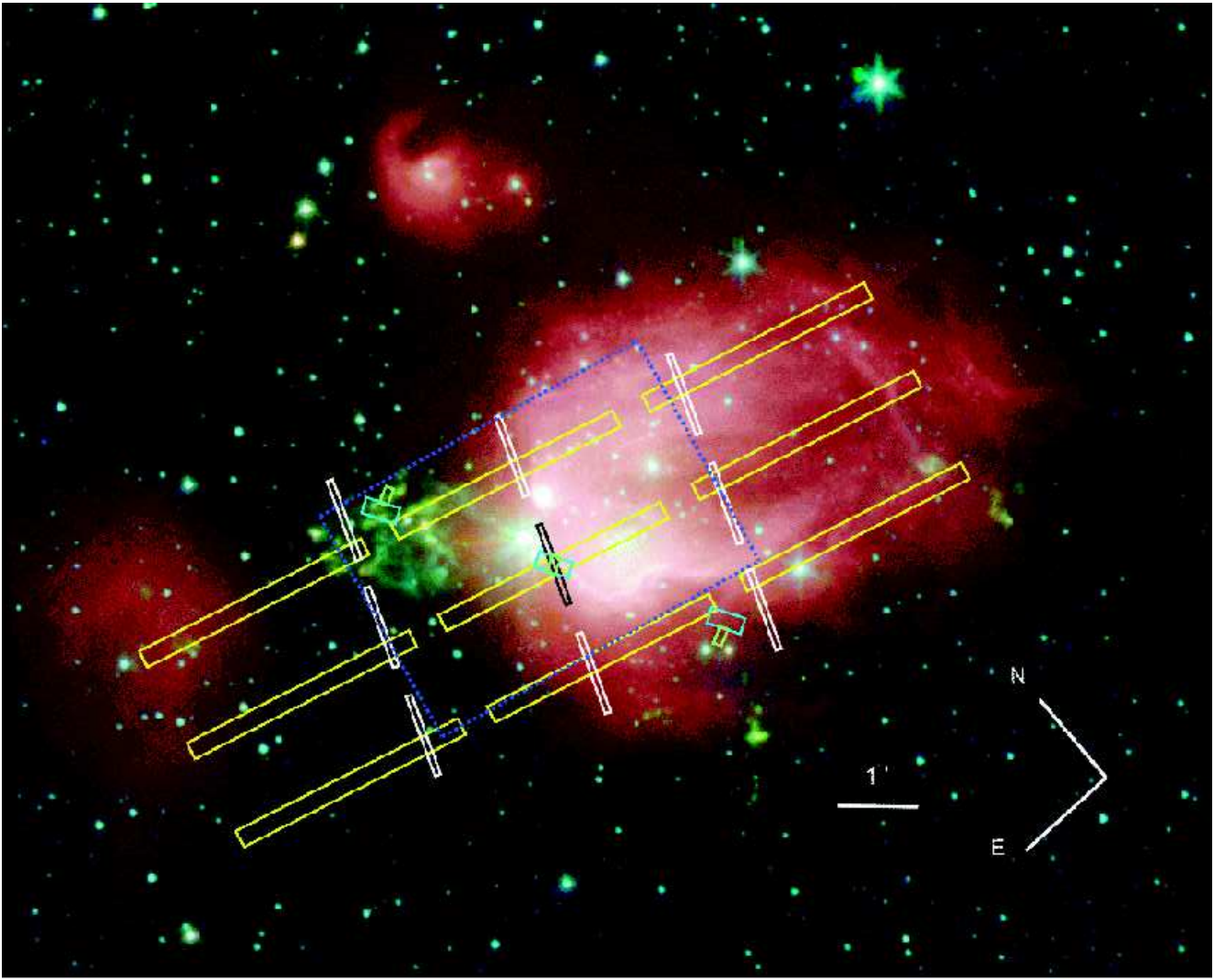}
\caption{Position of the IRS spectral slits on NGC~7129, 11 November 2003. 
The slits are Short Low (white or black), Long Low (yellow), Short 
High (green), and Long High (blue).  The $14' \times 12'$ image is an IRAC 
mosaic with the 3.6 (blue), 4.5 (green), 5.8 (orange), and 8.0 $\mu$m (red) 
filters (see Gutermuth et al. 2004).  The blue dotted line indicates
where IRS 16$\mu$m and ground-based 2.12 $\mu$m mapping have been obtained
(Fig.~\ref{cheap}). 
\label{rosebud} 
  }
\end{center}
\end{figure*}

\section{Introduction}\label{introduction}

NGC~7129 in Cepheus is a prototype young star-forming region and well-known 
reflection nebula, dominated by two young intermediate-mass ($M \simeq
3 - 10 M_\odot$) B stars, and the unusual Herbig Be star LkH$\alpha$~234.  
The winds from the early-type B stars, BD~+65$^\circ$1637 and BD~+65$^\circ$1638
(B2-B4; Strom et al.1972; Hillenbrand et al. 1992), but chiefly from the 
latter which is probably the older (Racine 1968),  have created a cavity of 
CO emission  near the center of the  nebula, and dense ridges of molecular 
material to the east, south, and west (Mitchell \& Matthews 1994; Fuente et al. 1998; 
Eisl\"{o}ffel 2000; Miskolczki et al. 2001).  Atomic and molecular lines observed
in an 80$''$ beam
with the Infrared Space Observatory Long Wavelength Spectrometer (LWS) 
indicate that the emission originates from photodominated excitations, possibly
in high density or clumpy environments as well (Lorenzetti et al. 1999; Giannini et al. 1999).
LkH$\alpha$~234 is estimated to be the youngest of these dominant stellar 
sources (Hillenbrand et al. 1992), probably formed in response to the compression of 
material on the northern edge (Mitchell \& Matthews), and is in a center of star 
forming activity that includes several pre-main sequence (PMS) objects, observed at 
millimeter wavelengths with high angular resolution (Fuente et al. 2001).  A low 
velocity ($\sim$10 km sec$^{-1}$) outflow extends over $\sim12'$ from LkH$\alpha$~234 
(Edwards \& Snell 1983), possibly as far as $\sim22'$ or some 8 pc as projected 
onto the sky (McGroarty, Ray, \& Bally 2004) at a distance of 1.25 kpc to NGC~7129
(Shevchenko \& Yakubov 1989), and an optical jet extends into the 
cavity (Ray et al. 1990).  The outflow and jet likely originate from a PMS object 
offset from LkH$\alpha$~234, as suggested by Cabrit et al. (1997) and then 
localized by Fuente et al. (2001) to a source $\sim3''.5$ to the NW.

In this paper we present the first mid-infrared spectra of two regions, one in the
outflow, and one in the ridge of H$_2$ emission near LkH$\alpha$~234, with the aim 
of exploring the excitation conditions in the environment surrounding this
young star in its early stages of evolution.  We estimate the intensity of the
UV radiation field from the strengths of the polyaromatic hydrocarbon (PAH)
band strengths, and determine the excitation mechanism responsible for 
H$_2$ line emission.  Imaging the nebula at 2.12~$\mu$m 
(H$_2$ 1 -- 0 S(1)) and 16~$\mu$m (continuum plus H$_2$  0 -- 0 S(1)) gives 
us a larger scale view on the spatial distribution of molecular gas.

\section{Spectroscopy and Imaging}

\subsection{IRS Spectroscopy}

NGC~7129 was observed on 11 November 2003 with the Infrared Spectrograph (IRS, 
described by Houck et al. 2004, this volume) onboard the {\em{Spitzer}}
Space Telescope (Werner et al. 2004a, this volume).   
A spectral map was performed with the slits centered on the 
H$_2$ ridge in the photodissocated region (PDR), at ${\rm{21^h43^m04.82^s}}$ 
$+66^{\circ}06'27''.7$ (all positions J2000.0), 29$''$.5 arcsec SW of 
LkH$\alpha$~234.  The map also includes a position in the strong molecular outflow, 
${\rm{21^h43^m19.8^s}}$ $+66^{\circ}07'50''.7$. 
Our analyses are concentrated on these two 
positions, nearest to LkH$\alpha$~234, and where the fullest wavelength coverage among 
the separate slits is most spatially confined (as described further below). All slit 
positions on the nebula are shown in Fig.~\ref{rosebud}, using the Early Release 
Observation\footnote{http://www.spitzer.caltech.edu/Media/releases/ssc2004-02/ssc2004-02a.shtml.
See also Gutermuth et al. (2004) and Muzerolle et al. (2004).}
of the nebula, obtained with the 
Infrared Array Camera (IRAC; Fazio et al. 2004, this issue) at 3.6, 
4.5, 5.8, and 8.0 $\mu$m.

Spectroscopy was obtained with each of the IRS modules, using total exposure 
times of 48 seconds with the Short Low module (SL, spanning 5.3 -- 14.5 $\mu$m at
spectral resolution $R = \lambda/\Delta\lambda \simeq$ 80 -- 120),  90 seconds 
with Short High (SH, 10 -- 19.7 $\mu$m, $R \simeq$ 700), and 120 seconds
with Long High (LH, 18.8 -- 37 $\mu$m, $R \simeq$ 700) at each slit position. Long
Low (LL, 14 -- 38 $\mu$m, $R \simeq$ 80 -- 120) data were also obtained, providing an 
internal check of the consistency with overlapping SH and LH data.  The relative 
spectrophotometry is consistent to within 10\% over the same sky area, and 
the LL data contain no new spectral information, rather degraded at the 
lower spectral resolution; therefore we present only the SH and LH
data here. The raw data have 
been processed in the SSC S9.1 pipeline to Basic 
Calibrated Data (BCD) products, and then extracted to 1-D spectra using the offline 
version of the post-BCD pipeline.  A region of $3''.6 \times 13''.5$ on the central map 
position was extracted from  the SL data at 10 $\mu$m. The full SH and LH 
slits were extracted and then scaled to the SL spectrum.  Two or three individual 
spectra over each wavelength range were obtained at each extracted position, and 
these were flux-calibrated, coadded, and sigma-clipped to produce a single spectrum
of the H$_2$ ridge spanning 5.3 -- 37 $\mu$m, and an outflow spectrum over
7.5 -- 37 $\mu$m.  Absolute fluxes are estimated to be uncertain by 
$\sim$20\% overall (Decin et al. 2004).

\begin{figure*}[th]
\begin{center}
\includegraphics[height=300pt,angle=0]{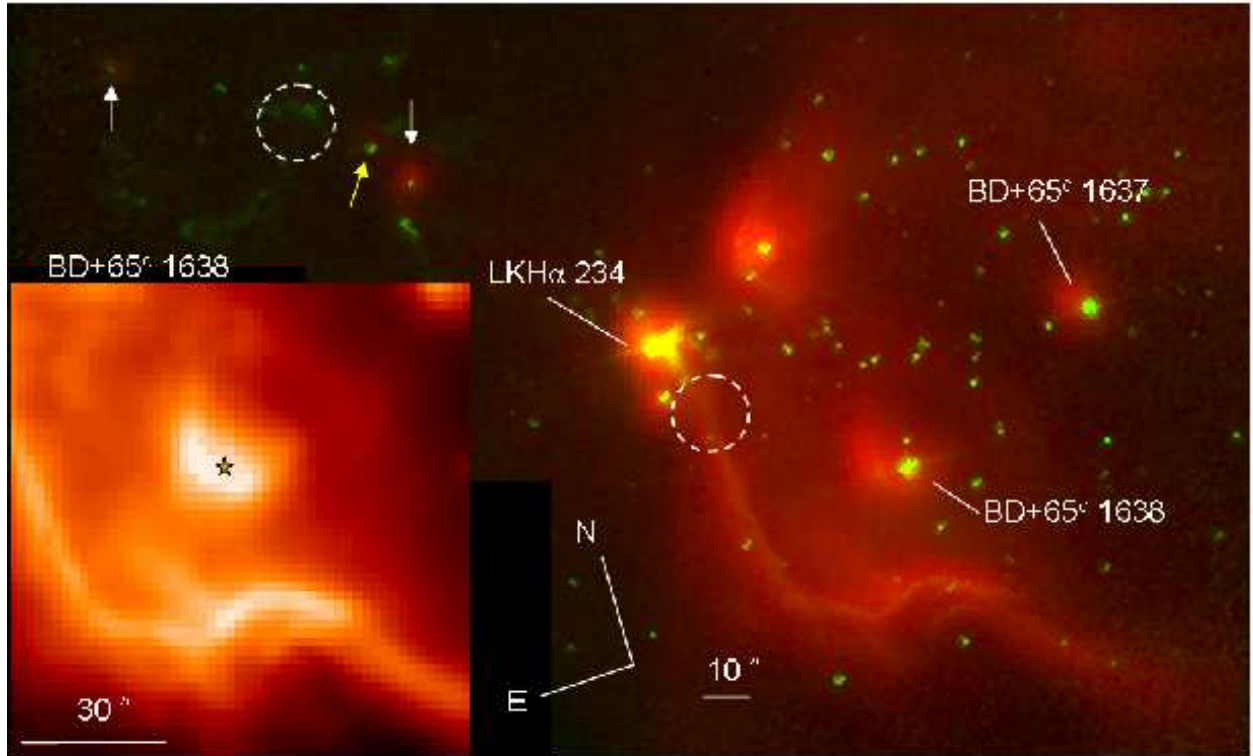}
\caption{IRS 16 $\mu$m (red) and CFHT 2.12 $\mu$m (green) mosaic of NGC~7129.
The bandpass of the 16$\mu$m camera covers
13.3 -- 18.7 $\mu$m.  Individual frames cover $54'' \times 82''$ with
a pixel scale of $1''.8$.  A mosaic of 25 positions, with 140 seconds of 
integration time per pixel on the sky, was obtained to cover an area of 
approximately $4'.3 \times 2'.5$. Regions of spectroscopic analysis are indicated by the circles with dashed lines.
Candidate PMS objects are indicated by the yellow and white arrows. A region of $\sim 80'' \times
85''$ around BD~+65$^\circ$1638 at 16$\mu$m is shown in the inset, to bring out details
of structured emission around the star, and the region of relative maximum surface brightness 
in the H$_2$ ridge.
\vspace{-2em}
\label{cheap} 
  }
\end{center}
\end{figure*}

\subsection{Ground-Based 2.12 $\mu$m and IRS 16 $\mu$m Imaging}

The H$_2$ 1 -- 0 S(1) 2.12$\mu$m data were obtained on 2003 November 10 -- 12 with
the 3.6m telescope at the Canada-France-Hawaii Telescope (CFHT) using 
the CFHTIR camera. The camera provides a $3'.6\times3'.6$
field of view and a scale of $0''.211$ pixel$^{-1}$.
The data were dark subtracted, flatfielded, and sky corrected in IRAF,
using the special near-IR data reduction package DIMSUM\footnote{The
Deep Infrared Mosaicing Software package of IRAF
scripts, developed by P. Eisenhardt, M. Dickinson, A. Stanford, and J. Ward, 
available at ftp://iraf.noao.edu/contrib/dimsumV2}.
The H$_2$ image has a total integration time of 18 minutes,
and been resampled to match the pixel scale of the 16$\mu$m~IRS image.

Mid-IR imaging of NGC~7129 was obtained with the IRS 16 $\mu$m camera on 10 October 
2003 and 29 November 2003.  Individual data frames were reduced in the S9.1 SSC 
pipeline to BCD products, and then registered based on reconstructed pointing,
which is accurate to $\sim$0.7$''$ (1-$\sigma$ radial).
The 16 $\mu$m and 2.12 $\mu$m mosaic is shown in Fig.~\ref{cheap}.
Several pixels are saturated at 16 $\mu$m by the three identified objects. 
In this field, Muzerolle et al. (2004, this issue) have proposed several
PMS objects on the basis of the colors extracted from IRAC and MIPS, three
of which are clearly evident by their 16 $\mu$m emission in Fig.~\ref{cheap}.
White arrows identify candidate Class 0/I objects, the yellow arrow 
corresponds to a Class II object.  At least one other PMS object in this
field is proposed by Muzerolle et al., but is not evident by a 16 $\mu$m
thermal excess.

Fig.~\ref{cheap} also shows an arc of 16 $\mu$m emission in the vicinity of 
BD~+65$^\circ$1638, almost concentric with the eastern H$_2$ ridge.  This structure is
not discerned in IRAC or MIPS imaging (cf. Gutermuth et al. 2004 and Muzerolle
et al. 2004), possibly as a consequence of the temperature and 
surface brightness of the material.  At moderate angular resolution ($1'$), 
Matthews et al. (2003) identify a dense concentration of neutral hydrogen
(21 cm) coincident with BD~+65$^\circ$1638, as well as an extensive region
of photodissociated H$_2$ accompanied by a small H~{\sc{ii}} region around
this star.  Miskolczi et al. (2001) have proposed that the cavity of molecular
material around BD~+65$^\circ$1638 
has been created by photodissociation of H$_2$ into neutral hydrogen, 
and that the gas and radiation pressure from within the cavity are responsible 
for sweeping material into a shell that appears as the H$_2$ ridge. 
The presence of this concentrated structure (acknowledging some saturation
of an area of radius $\sim2''.5$ centered on BD~+65$^\circ$1638) clearly shows that 
material, either molecular gas or dust, in not entirely evacuated from this 
region.  Further spectroscopy is needed determine its properties. 

\begin{figure*}[ht]
\begin{center}
\includegraphics[height=500pt,angle=90]{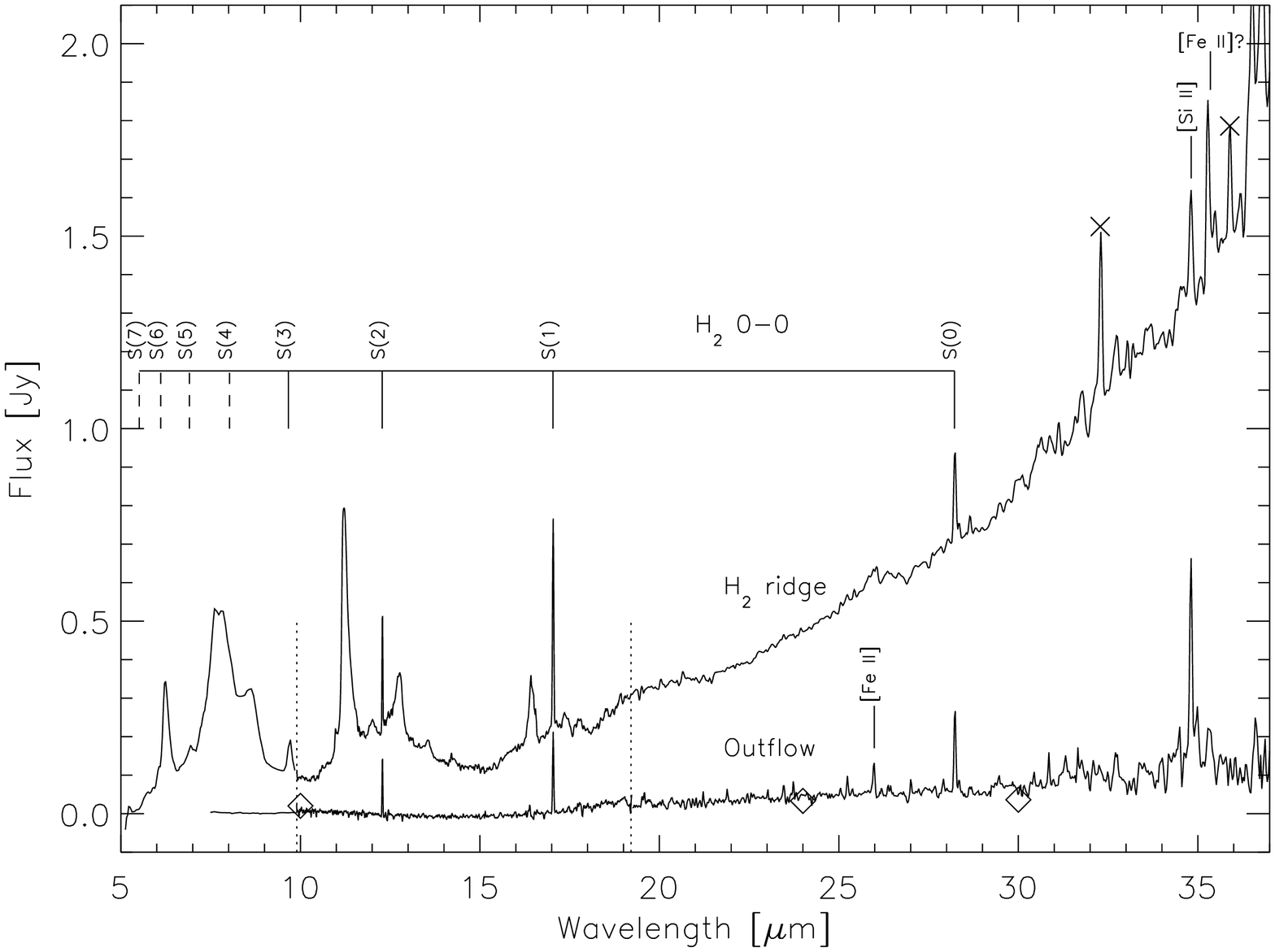}
\caption{IRS spectra of the outflow and PDR near LkH$\alpha$~234, as indicated.
Data are on their natural flux scales, without offsetting.  Diamonds represent
emission levels of the sky (ISM and zodiacal light) background.  The atomic nebular
and H$_2$ rotational are identified, where dashed lines indicate where H$_2$
lines in the SL range are severely blended with surrounding PAH emission.  The
dotted lines show where the SL, SH, and LH data have been joined. Cosmic
ray artifacts are indicated by ''$\times$''. 
\label{spec} 
  }
\end{center}
\end{figure*}

\begin{table}[th!]
\caption{Rotational H$_2$ 0 -- 0 observed fluxes, column densities, and excitation
temperatures\tablenotemark{a}. 
\label{lines}}
\begin{tabular}{l|lrr|lrr}
\tableline\tableline
\multicolumn{1}{c}{Line}  & \multicolumn{3}{c}{PDR}  & \multicolumn{3}{c}{Outflow} \\
\multicolumn{1}{c}{[$\mu$m]} & \multicolumn{1}{|c}{Flux} & $\ln(N/g)$ & $E_{\rm{up}}/k$ & \multicolumn{1}{c}{Flux} & $\ln(N/g)$ & $E_{\rm{up}}/k$ \\
\tableline
S(6) 6.11 &  61.0 (1.3)\tablenotemark{b} & \multicolumn{1}{c}{\nodata} & \multicolumn{1}{c|}{\nodata} & \multicolumn{1}{c}{\nodata} & \multicolumn{1}{c}{\nodata} & \multicolumn{1}{c}{\nodata} \\
S(5) 6.91 &  23.0 (1.7)\tablenotemark{c} & 38.55 & 4415 & \multicolumn{1}{c}{\nodata} & \multicolumn{1}{c}{\nodata} & \multicolumn{1}{c}{\nodata} \\
S(4) 8.02 &  41.0 (2.5)\tablenotemark{b} & \multicolumn{1}{c}{\nodata}  & \multicolumn{1}{c|}{\nodata} & \multicolumn{1}{c}{\nodata} & \multicolumn{1}{c}{\nodata} & \multicolumn{1}{c}{\nodata} \\
S(3) 9.66 &  8.75 (0.77) & 40.01 & 2333 &  \multicolumn{1}{c}{\nodata} & \multicolumn{1}{c}{\nodata} & \multicolumn{1}{c}{\nodata} \\
S(2) 12.28 & 1.20 (0.03) & 40.84 & 1682 &   0.54 (0.02) & 40.04 & 1682 \\
S(1) 17.03 & 1.70 (0.04) & 42.41 &  844 &   0.54 (0.02) & 41.27 &  844 \\
S(0) 28.22 & 0.37 (0.02) & 45.60 &  510 &   0.33 (0.01) & 45.49 & 510 \\
\tableline
           &             & $N_{\rm{tot}}$ & $T_{\rm{ex}}$ &  & $N_{\rm{tot}}$ & $T_{\rm{ex}}$  \\
           &             & 1.17 (0.04) & 658 (9)  &  & 13.4 (0.7) & 248 (12) \\
\tableline
\end{tabular}
\tablenotetext{a}{Fluxes in units of 10$^{-11}$ W cm$^{-2}$ sr$^{-1}$, uncertainties
are in parentheses.  Excitation temperature $T_{\rm{ex}}$ is in K, and
total column density density $N_{\rm{tot}}$ is in 10$^{19}$ cm$^{-2}$.}
\tablenotetext{b}{Severely blended with neighboring PAH band.}
\tablenotetext{c}{Blended with neighboring PAH band.}
\end{table}

\section{Spectroscopic Results}

The IRS spectra of the PDR/H$_2$ ridge and outflow are presented in Fig.~\ref{spec},
as labeled.  The spectra are not offset from each other. Sky background levels
indicated by the triangles at 10, 24, and 30 $\mu$m are estimated from a 
{\em{Spitzer}}-centric, COBE/DIRBE-based\footnote{Avaialable in the SPOT 
observations planning tool.} model of the thermal emission by 
zodiacal and ISM dust in the line of sight to NGC~7129, on its date of observation.
At these levels, we surmise that the outflow contains negligible amounts
of warm dust. 


The PDR spectrum exhibits the usual set of PAH bands, at peak positions of
6.22, 7.63, 8.63, 11.22, 12.75, and 16.46 $\mu$m, dominated by the 7.63 and
11.22 $\mu$m bands.  Secondary peaks that may be associated with PAH emission
are present at 11.00, 11.99, 13.55, 14.24, and possibly 17.37 $\mu$m.  These
features are not previously identified, to our knowledge, and Werner et al.
(2004b, this volume) identify other new subsidiary bands in IRS spectra of 
the NGC~7023 reflection nebula, including the 17.4 $\mu$m feature. 

The strength of the PAH band emission allows us to estimate the intensity of the
UV radiation field in the vicinity of LkH$\alpha$~234.  Using a two-component
blackbody with temperatures 75~K and 320~K and unity emissivity to approximate
the thermal continuum, we measure strengths of the 6.22, 7.63, 11.22, and 
12.75 $\mu$m bands in the range of $6.0 - 30 \times 10^{-5}$
W~m$^{-1}$~sr$^{-1}$.   The observational trend of PAH band intensities plotted
against UV radiation field density by Boulanger et al. (1998) for a set of
PDRs implies a field strength of 800 to 900 times the solar value.  This
is close to the value of NGC~7023, which is measured by Boulanger et al.
from a region of maximum emission in the PDR, in proximity
to the PMS Herbig B2-B3 star HD~200775 which is similar to our PDR location 
near  BD+65$^{\circ}$~1638, without consideration in either case for
shielding by circumstellar material.  And like NGC~7023, the PAH bands
are lower in intensity at this UV field strength than given by 
a simple power law extrapolation from PDRs with lower field strengths.  
This could be due to a lower abundance of small dust grains in high radiation 
fields, as suggested by Boulanger et al.


The PDR and outflow spectra both exhibit pure rotational lines of H$_2$, 
arising from optically thin quadrupole transitions that are readily 
thermalized at moderate volume densities.  This allows us to estimate
H$_2$ column densities and excitation temperatures under the assumption
of LTE (e.g., Gredel 1994), in order to gain insight into the excitation conditions of the
line forming regions.  We follow excitation diagram methods developed by
Burton (1992) and Gredel (1994), and applied to comparable environments (as in, 
e.g., Cepheus~E by Moro-Martin et al. 2001). The column density for
transition $j$ is $N_j$ = 4$\pi F_j / \Omega E_j g_j A_j$, where $E_j$ is the
energy of the upper level, $g_j$ is the statistical weight of ortho- and
para-transitions, and $A_j$ is the transition probability. The
line column density is related to the total column density and excitation
temperature  $T_{\rm{ex}}$ by $\ln(N_j/g_j) = ln(N_{\rm{tot}}/Q)-E_j/(kT_{\rm{ex}})$,
and may be solved for with a Boltzmann excitation diagram.  
The partition function $Q$ was determined with an ortho-para ratio of 3 under
the assumption of thermal equilibrium (Moro-Martin et al. 2001).
For the observed line fluxes per unit solid angle $F_j/\Omega$, we
we have extracted over regions of 40~arcsec$^2$ and 62~arcsec$^2$ in the
5.3 -- 14 $\mu$m and 14 -- 35 $\mu$m ranges, respectively.  In these areas we
must view our measurements as spatially integrated.
The computed values for ln(Nj/gj), Eup/K along with
the final estimates of the total H2 column density and excitation
temperature for the PDR and outflow are listed in Table 1.  The observed line 
fluxes per unit solid angle $F_j/\Omega$, 1-$\sigma$ measurement uncertainties, and 
computed values of $\ln(N_j/g_j)$, 
$E_{\rm{up}}/k$ along with final estimates of the total H$_2$ column density and 
excitation temperature for the PDR and outflow are listed in Table~\ref{lines}.
The lines were measured with the line-fitting utilities in 
the ISAP\footnote{ISO Spectral Analysis Package: \\ http://www.ipac.caltech.edu/iso/isap/isap.html}, 
which allows multiple fits of Lorentzian and Gaussian profiles for deblending, suited
to the PAH features (e.g., Boulanger et al. 1998) and the unresolved H$_2$ lines.  
This permits the H$_2$ S(4) and S(5) fluxes to be separated from the
7.6 and 8.6~$\mu$m PAH bands.
The PDR line intensities are corrected for reddening using the Draine \& Lee (1984)
extinction law, and $A_V$ = 2.6 mag estimated by Hamann \& Persson (1992) 
towards LkH$\alpha$~234.  The extinction across the nebula probably varies
widely, and the ISM component is likely to be much lower along the line of sight
to our H$_2$ ridge observation, but the differences to the S(0) -- S(2) lines will
be small ($<$ 5\%), and measurement uncertainties of the S(3) -- S(5) lines arise 
predominantly from the blending with the PAH features and the flux calibration. 
The S(6)/S(5) or S(4)/S(5) line ratios could be also affected by an ortho-para
ratio different than the canonical (thermal equilibrium) value of 3, as in the 
presence of shocks  (Wilgenbus et al. 2000), or by the superposition of 
multiple layers along the line of sight with different physical conditions. 
From this standpoint we must view our spatially-limited measurements as mean 
indicators of the physical conditions in both the PDR and outflow.

The factor $\sim$2.5 in $T_{\rm{ex}}$ between the PDR and the outflow seems
quite reasonable, for line emission in the outflow arising from collisional
(dissociative J-shock) excitation.  We favor the J-shock mechanism over
C-shocks in the outflow, based on the S(0) line intensity and strong 
[Si~{\sc{ii}}] 34.815 $\mu$m emission.  In the outflow the [Si~{\sc{ii}}] 
and [Fe~{\sc{ii}}] line fluxes are 1.06 ($\pm$0.15) $\times 10^{-20}$ and
2.36 ($\pm$0.38) $\times 10^{-21}$ W~cm$^{-2}$, respectively.  In the PDR, the
[Si~{\sc{ii}}] line flux is 7.92 ($\pm$0.13) $\times 10^{-21}$ W cm$^{-2}$. 
Interestingly, the measured line intensities of the S(0) line in the PDR and outflow 
are quite similar, which might be explained by a photodissociating 
component in the outflow.  As for velocities, Edwards \& Snell (1983)
measured a maximum CO expansion of 10 km sec$^{-1}$, associated with the
outflow.  In J-shock models (Hollanbach \& McKee 1989) for a faster jet-related 
component, H$_2$ lines are not a good diagnostics of shock
velocities, as their intensities are almost flat as function of velocity at 
the densities we expect ($\sim 10^5$ cm$^{-3}$).  But we know from the
absence of [Ne~{\sc{ii}}] and presence of H$_2$ that the velocity must be 
less than 50 km sec$^{-1}$.  At the densities and low $T_{\rm{ex}}$ we
measure, we estimate J-shock velocities in the range of 10 -- 30 km sec$^{-1}$.

The H$_2$ spectrum (and thus the $T_{\rm{ex}}$ we infer) for the H$_2$ ridge 
can be explained by the standard Draine \& Bertoldi (1996)
models for a stationary PDR, and are similar to conditions in the
S~140 PDR (Timmermann et al. 1996).  This does not rule out contribution
from non-radiative processes that are reasonable in this environment, where
either or both of BD~+65$^\circ$1638 and BD~+65$^\circ$1637 are sweeping material 
outwards by mechanical wind momentum, to form some H$_2$ by collisional excitation.  
Indeed far infrared spectroscopic observations of the LKH$\alpha$~234 region with
multiple pointings using the ISO-LWS (Lorenzetti et
al. 1999) already indicate that the PDR is the dominant component for the 
H$_2$ fluorescent emission, but the presence of C-shocked material
cannot be ruled out.  Our observations lack the spatial information with additional
pointings to make a comparison with a more complex, clumped PDR model 
(e.g., Burton et al. 1990) as done by Lorenzetti et al. (1999; see their Fig. 6).
It is anyway useful to note that collisional excitation, if present in a clumped
environment, cannot be done in cooling zones behind dissociative J-shocks 
(cf. Hollenbach \& McKee 1989), where $T_{\rm{ex}}$ is lower than 300 -- 400 K 
at our densities, but more likely in C-shocks with pre-shock
H$_2$ volume density, magnetic field, ionization fraction, and shock speed
conditions found, for example, in GGD37 in Cepheus A West (Wright et al. 1996).  This region 
is also a site of intermediate mass star formation with a massive outflow, and 
bow-shock like clumps and filaments of H$_2$ emission (Hartigan et al. 1996),
exhibiting very comparable H$_2$ rotational line fluxes 
(and $T_{\rm{ex}} \simeq$ 700 K). Aperture average column densities are one order 
of magnitude higher than in the NGC~7129 H$_2$ ridge, which gives us a 
correspondingly lower H$_2$ mass of $\sim 10^{-3} M_\odot$.

\section{Summary and Conclusions}

With the IRS we have obtained the first mid-IR spectroscopy
of regions around LkH$\alpha$~234, in the recognizable outflow and in 
a nearby ridge of H$_2$ emission.   The measured intensities of the
PAH bands in the H$_2$ ridge allowed us to infer the strength of the
UV radiation field, which is comparable to that estimated by
Boulanger et al. (1998) in NGC~7023.  The rotational H$_2$ emission line 
ratios infer an (LTE) aperture average excitation temperature of 657~K and 
column density of 1.2 $\times 10^{19}$ cm$^{-2}$, quite similar to
the warm-H$_2$ regions of Cep~A that are well matched by formation in 
non-dissociative C-shocks.   
The cooler H$_2$ emission in the outflow is consistent with formation by 
collisional excitation in J-shocks, with velocities of 10 -- 30 km~sec$^{-1}$.
A photodissociating component is suggested by similarity of S(0) line intensities 
in the outflow and PDR/H$_2$ ridge.  There is no significant contribution of
thermal emission from warm dust in the outflow.

The origin of the PDR cannot be localized with certainty from the IRS observations
presented here.  Lorenzetti et al. (1999) presented ISO-LWS observations of
[C~{\sc{ii}}] 158 $\mu$m, [O~{\sc{i}}] 63 $\mu$m, and [O~{\sc{i}}] 146 $\mu$m
emission whose spatial morphology and derived densities and UV field strength 
across the nebula favor BD+65$^{\circ}$~1637, arguing that this is consistent
with observations by Fuente et al. (1998) of a $^{13}$CO cavity created by
the young star.  Miskolczi et al. (2001) have subsequently shown with $^{12}$CO 
and $^{13}$CO maps that the cavity is instead centered on the older star 
BD+65$^{\circ}$~1638, which Matthews et al. (2003) recently find in 21 cm
maps to be surrounded by an extensive photodissociated H$_2$ and a small 
H~{\sc{ii}} region.  On the other hand, our detection of warm material
at 16 $\mu$m around BD+65$^{\circ}$~1638 clearly shows that this region is
not free of material, and though BD+65$^{\circ}$~1638 is not an embedded object,
further spectroscopic observations are needed to ascertain its nature.

\begin{acknowledgements}
We thank Louise Edwards and the CFHT staff for their help with the
CFHTIR observations.  We also thank an anonymous referee for helpful comments
on the manuscript. 
\end{acknowledgements}






\end{document}